\documentclass[12pt]{iopart}
\usepackage{graphicx}
\usepackage{cite}
\usepackage{amssymb}
\newcommand{\be}{\begin{equation}}
\newcommand{\ee}{\end{equation}}
\newcommand{\bel}[1]{\begin{equation}\label{#1}}
\newcommand{\bea}{\begin{eqnarray}}
\newcommand{\eea}{\end{eqnarray}}
\newcommand{\ba}{\begin{array}}
\newcommand{\ea}{\end{array}}

\newcommand{\bra}[1]{\mbox{$\langle \, {#1}\, |$}}
\newcommand{\ket}[1]{\mbox{$| \, {#1}\, \rangle$}}
\newcommand{\exval}[1]{\mbox{$\langle \, {#1}\, \rangle$}}

\renewcommand{\eta}{n}

\begin{document}

\title{Zero-range process with open boundaries}
\author{E. Levine$^{(a)}$, D. Mukamel$^{(a)}$, and G.M. Sch\"{u}tz$^{(b)}$}
\address{$(a)$ Department of Physics of Complex Systems,
Weizmann Institute of Science, Rehovot, Israel 76100.\\
$(b)$ Institut f\"{u}r Festk\"{o}rperforschung, Forschungszentrum
J\"{u}lich, 52425 J\"{u}lich, Germany.}

\date{\today}

\begin{abstract}
We calculate the exact stationary distribution of the
one-dimensional zero-range process with open boundaries for
arbitrary bulk and boundary hopping rates. When such a
distribution exists, the steady state has no correlations between
sites and is uniquely characterized by a space-dependent fugacity
which is a function of the boundary rates and the hopping
asymmetry. For strong boundary drive the system has no stationary
distribution. In systems which on a ring geometry allow for a
condensation transition, a condensate develops at one or both
boundary sites. On all other sites the particle distribution
approaches a product measure with the finite critical density
$\rho_c$. In systems which do not support condensation on a ring,
strong boundary drive leads to a condensate at the boundary.
However, in this case the local particle density in the interior
exhibits a complex algebraic growth in time. We calculate the bulk
and boundary growth exponents as a function of the system
parameters.
\end{abstract}

\maketitle

\section{Introduction}

The zero-range process (ZRP) has originally been introduced in
1970 by Spitzer \cite{Spit70} as a model system for interacting
random walks, where particles on a lattice hop randomly to other
neighboring sites. The hopping rates $w_n$ depend only on the
number of particles $n$ at the departure site. This model has
received renewed attention because of the occurrence of a
condensation transition \cite{Evan96,Krug96,Oloa98,Evan00,Jeon00a}
analogous to Bose-Einstein condensation and because of its close
relationship with exclusion processes \cite{Schu03}. Condensation
phenomena are well-known in colloidal and granular systems (see
\cite{Shim04} for a recent study making a connection with the
zero-range process), but appear also in other contexts, such as
socio-economics \cite{Burd01} and biological systems \cite{Froh75}
as well as in traffic flow \cite{traffic} and network theory
\cite{Bian01,Doro03}. In the mapping of the ZRP to exclusion
processes in one space dimension, condensation corresponds to
phase separation. The ZRP has served for deriving a quantitative
criterion for the existence of non-equilibrium phase separation
\cite{Kafr02} in the otherwise not yet well-understood driven
diffusive systems with two conservation laws. For the occurrence
of condensation the dimensionality of the ZRP does not play a role
and we shall consider only the one-dimensional (1d) case.

Most studies of the ZRP focus on periodic boundary conditions or
the infinite system. Under certain conditions on the rates $w_n$
(see below) the grand-canonical stationary distribution is a
product measure, i.e. there are no correlations between different
sites \cite{Andj82}. In addition, an exact coarse-grained
description of the {\it dynamics} is possible in this case in
terms of a hydrodynamic equation for the particle density
$\rho(x,t)$ \cite{Kipn99,Gros03b}.

In zero-range processes for which the hopping rates $w_n$ admit
condensation, one finds that above a critical density $\rho_c$ a
finite fraction of all particles in the system accumulate at a
randomly selected site, whereas all other sites have an average
density $\rho_c$ \cite{Evan00,Jeon00b,Gros03a}. The large scale
dynamics of condensation has been studied in terms of a coarsening
process \cite{Gros03a,Godr03}. The steady-state and the dynamics
of open systems which may admit condensation has not been
addressed so far.

In the present work we consider a ZRP on an open chain with
arbitrary hopping rates and boundary parameters. Particles are
added and removed through the boundaries. In the interior of the
system hopping may either be symmetric or biased in one direction.
We calculate the exact steady-state distribution, when it exists,
and find it to be a product measure. In order to study
condensation phenomena in the open system we analyze in detail a
particular but generic ZRP which admits condensation. In this
model the hopping rates for large $n$ take the form $w_n=1+b/n$.
On a periodic lattice it is known that a condensation takes place
at high densities when $b>2$. We find that in an open system and
for a weak boundary drive the model evolves to a non-critical
steady-state. On the other hand, if the boundary drive is
sufficiently strong, the system may develop a condensate on one or
both of its boundary sites even for $b<2$. The number of particles
in the condensate grows linearly in time due to the strong
boundary drive. In this case the interior of the system may either
(a) reach a sub-critical steady-state, (b) reach a critical
steady-state, or (c) it may evolve such that the local particle
density exhibits a complex algebraic growth in time. Which
behavior is actually realized depends on the boundary rates, on
the parameter $b$, and on the asymmetry in the hopping rates.

The paper is organized as follows. In Sec.~\ref{sec:def} we define
the ZRP with open boundaries, and give some examples of possible
hopping rates $w_n$. The exact steady-state distribution for
general boundary parameters and rates $w_n$ is derived in
Sec.~\ref{sec:st}. In Sec.~\ref{sec:finite} we consider the
behavior of finite systems for totally asymmetric, partially
asymmetric and symmetric hopping rates. The long-time temporal
behavior of local densities is obtained, and supporting numerical
simulations are presented. The long-time behavior of bulk
densities in an infinite system is exactly obtained using a
hydrodynamical approach in Sec.~\ref{sec:hydro}. Finally, we
present our summary and conclusions in Sec.~\ref{sec:summary}.

\section{Zero-range processes with open boundaries}
\label{sec:def}

\begin{figure}
\centerline{\includegraphics[scale=0.45]{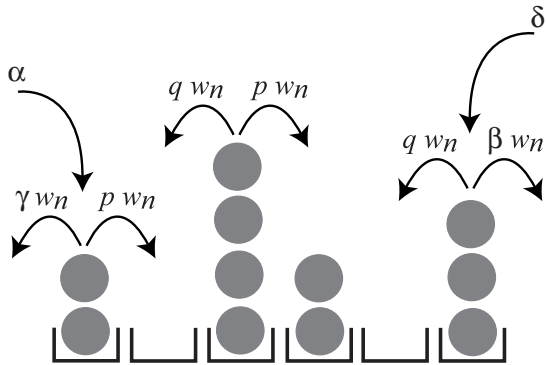}}
\caption{Graphic representation of the ZRP model with open
boundaries.} \label{fig:toy}
\end{figure}

The ZRP on an open 1d lattice with $L$ sites is defined as
follows. Each site $k$ may be occupied by an arbitrary number $n$
of particles. In the bulk a particle at site $k$ (say, the topmost
of $n$ particles) hops randomly (with exponentially distributed
waiting time) with rate $p w_n$ to the right and with rate $qw_n$
to the left. Without loss of generality, we take throughout this
paper $p \geq q$, so that particles in the bulk are driven to the
right. At the boundaries these rules are modified. At site $1$ a
particle is injected with rate $\alpha$, hops to the right with
rate $pw_n$, and is removed with rate $\gamma w_n$. At site $L$ a
particle is injected with rate $\delta$, hops to the left with
rate $qw_n$, and is removed with rate $\beta w_n$ (see
figure~\ref{fig:toy}).

Examples for such processes include the trivial case of
non-interacting particles ($w_n=n$) and the pure chipping process
($w_n=1$) \cite{chip}. In the mapping to exclusion processes where
the particle occupation number becomes the interparticle distance,
the chipping model defined on a ring maps onto the simple
exclusion process which is integrable and can be solved by the
Bethe ansatz, combinatorial methods, recursion relations and
matrix product methods \cite{Derr98,Schu00}. A largely unexplored
but interesting integrable model with rates $w_n = (1-q^n)/(1-q)$
that interpolates between these two cases has been found in
\cite{Povo04}. The truncated chipping process with $w_n=1$ for $1
\leq n \leq K$ and $w_n = \infty$ for $n > K$ maps onto the
drop-push model \cite{Schu96} which is also integrable
\cite{Alim98}. For $w_1=w$ and $w_n = 1$ for $n\geq 2$ one obtains
the version of the non-integrable KLS model \cite{Katz84} that has
been introduced in \cite{Anta00} as a toy model for traffic flow
with a nonsymmetric current density relation. A parallel updating
version of this model was found to correspond to a broader class
of traffic models \cite{Levi04T} as well as an integrable family
of ZRP's \cite{Povo04b} which includes the asymmetric avalanche
process \cite{Prie01}.

In a similar mapping to exclusion processes (but in a continuum
limit) the family of models with hopping rate \bel{rodmodel} w_n =
\left[1+ \frac{b'}{n}\right]^{-1} \ee describes the diffusion of
interacting rods on the real line \cite{Scho04}. The condensation
model \cite{Oloa98,Evan00,Kafr02,Jeon00b,Gros03a,Godr03} with
\bel{condmodel} w_n = 1+ \frac{b}{n} \ee is a generic model that
exhibits the condensation phenomenon described in the introduction
for $b>2$. Both models are non-integrable and hence alternative
tools must be employed for deriving information about their
dynamical behavior. We remind the reader that this model is
generic in the sense that it represents the complete family of
models with rates of the form $w_n = 1 + b/n + O(n^{-s})$ where
$s>1$. Other ZRP's which exhibit condensation are defined by the
rates $w_n = 1+b/n^\sigma$ with $0<\sigma<1$ and $b>0$
\cite{Kafr02} or by rates approaching zero \cite{Jeon00a}.

\section{Stationary distribution}
\label{sec:st}

A state of the model at time $t$ may be defined through a
probability measure $P_\eta$ on the set of all configurations
$\eta=(\eta_1,\eta_2, \dots, \eta_L)$, $\eta_k\in\mathbb{N}$. Here
$\eta_k$ is the number of particles on site $k$. To calculate the
stationary distribution it is convenient to represent the
generator $H$ of this process in terms of the quantum Hamiltonian
formalism \cite{Schu00} where one assigns a basis vector
$|\eta\rangle$ of the vector space $(\mathbb{C}^\infty)^{\otimes
L}$ to each configuration and the probability vector is defined by
$|P\rangle=\sum_\eta P_\eta |\eta\rangle$. It is normalized such
that $\langle s|P\rangle=1$ where $\langle s|=\sum_\eta
\langle\eta|$ and $\bra{\eta}\eta'\,\rangle =
\delta_{\eta,\eta'}$. The time evolution described above is given
by the master equation
\begin{equation}
  \label{master}
  \frac{d}{dt} |P(t)\rangle = -H |P(t)\rangle
\end{equation}
through the ``quantum Hamiltonian'' $H$. This operator has off-diagonal
matrix elements $H_{\eta,\eta'}$ which are
the hopping rates between configurations $\eta,\eta'$
and complementary diagonal elements to preserve conservation of probability.

Since we have only nearest neighbor exchange processes the
Hamiltonian in (\ref{master}) can be written as
\begin{equation}
  \label{H}
H = h_1 + h_L +\sum_{k=1}^{L-1} h_{k,k+1},
\end{equation}
where $h_{k,k+1}$ acts nontrivially only on sites $k$ and $k+1$
(corresponding to hopping) while $h_1,h_L$ generates the boundary
processes specified above. For the ZRP we define the
infinite-dimensional particle creation and annihilation matrices
\bel{ladder} a^+ = \left(\begin{array}{ccccc}
    0 & 0 & 0 & 0 & \dots \\
    1 & 0 & 0 & 0 & \dots \\
    0 & 1 & 0 & 0 & \dots \\
    0 & 0 & 1 & 0 & \dots \\
    \dots & \dots & \dots & \dots & \dots \\
  \end{array}\right), \quad
a^- = \left(\begin{array}{ccccc}
    0 & w_1 & 0 & 0 & \dots \\
    0 & 0 & w_2 & 0 & \dots \\
    0 & 0 & 0 & w_3 & \dots \\
    0 & 0 & 0 & 0 & \dots \\
    \dots & \dots & \dots & \dots & \dots \\
  \end{array}\right)
\ee
as well as the diagonal matrix $d$ with elements
$d_{i,j} = w_i \delta_{i,j}$. With these matrices we have
\bel{hoppingpart}
- h_{k,k+1} = p (a^-_{k}a^+_{k+1} - d_k) + q (a^+_{k}a^-_{k+1} - d_{k+1})
\ee
and
\bel{boundarypart}
- h_1 = \alpha (a^+_1 - 1) + \gamma (a^-_1-d_1), \quad
- h_L = \delta (a^+_L - 1) + \beta (a^-_L-d_L).
\ee
The ``ground state'' of $H$ has eigenvalue 0. The corresponding right
eigenvector is the stationary distribution which we wish to calculate.

Guided by the grand-canonical stationary distribution of the
periodic system we consider the grand-canonical single-site
particle distribution where the probability to find $n$ particles
on site $k$ is given by
\bel{marginal} P^\ast(\eta_k=n) =
\frac{z_k^n}{Z_k} \prod_{i=1}^n w_i^{-1}.
\ee
Here the empty
product $n=0$ is defined to be equal to 1 and $Z_k$ is the local
analogue of the grand-canonical partition function
\bel{Z} Z_k
\equiv Z(z_k) = \sum_{n=0}^\infty z_k^n \prod_{i=1}^n w_i^{-1}.
\ee
The corresponding probability vector $|P^\ast_k)$ with the
components $P^\ast(\eta_k=n)$ satisfies
\bel{coherent} a^+
|P^\ast_k) = z_k^{-1} d |P^\ast_k), \quad a^- |P^\ast_k) = z_k
|P^\ast_k). \ee
The proof of this property is by straightforward
calculation.

As an ansatz for calculating the stationary distribution we take
the $L$-site product measure with the one-site marginals (\ref{marginal})
which is given by the tensor product
\bel{product}
\ket{P^\ast} = |P^\ast_1) \otimes |P^\ast_2) \otimes \dots \otimes |P^\ast_L)
\ee
and according to (\ref{coherent}) satisfies
\bea
- H \ket{P^\ast} & = & \left[\sum_{k=1}^{L-1} (pz_k-qz_{k+1})
(z_{k+1}^{-1} d_{k+1} - z_k^{-1} d_k) \right.\nonumber \\
& & + \left.(\alpha-\gamma z_1) (z_{1}^{-1} d_{1} - 1) + (\delta -
\beta z_L) (z_{L}^{-1} d_{L} - 1)\right]\ket{P^\ast}. \eea The
uncorrelated particle distribution (\ref{product}) is stationary
if and only if all terms on the right hand side of this equation
cancel. It is not difficult to see that this is satisfied for the
following stationarity conditions on the fugacities $z_k$ \be p
z_k - q z_{k+1} = \alpha - \gamma z_1 =   \beta z_L -\delta \equiv
c\;. \ee The quantity $c$ is the stationary current.

This recursion relation has the unique solution \bel{steadystate}
z_k = \frac{\left[(\alpha+\delta)(p-q) - \alpha\beta +
\gamma\delta\right] \left(\frac{p}{q}\right)^{k-1} - \gamma\delta
+ \alpha\beta \left(\frac{p}{q}\right)^{L-1}}{\gamma(p-q-\beta) +
\beta(p-q+\gamma) \left(\frac{p}{q}\right)^{L-1}} \;,\ee and the
current is given by \bel{current} c = (p-q) \frac{- \gamma\delta +
\alpha\beta \left(\frac{p}{q}\right)^{L-1}}{\gamma(p-q-\beta) +
\beta(p-q+\gamma) \left(\frac{p}{q}\right)^{L-1}}\;. \ee Thus,
given the rates $w_n$ the stationary distribution is unique and
completely specified by the hopping asymmetry and the boundary
parameters. The stationary density profile follows from the
fugacity profile through the standard relation \be \rho_k = z_k
\frac{\partial}{\partial z_k} \ln{Z_k}\;, \ee where $Z_k = Z(z_k)$
is determined by the bulk hopping rules, as given in (\ref{Z}).

We remark that up to corrections exponentially small in the system
size $L$, the bulk fugacity is a constant \bel{zeff}
z_{\mathrm{eff}} = \frac{\alpha}{p-q + \gamma}\;, \ee
 that depends only on the
{\it left} boundary rates. This is in agreement with the observation
\cite{Gros03b} for special boundary parameters and can be
explained in terms of the more general theory of boundary-induced
phase transitions \cite{Kolo98}.

Some special cases of (\ref{steadystate}) deserve mentioning.\\
(1) Setting the boundary extraction rates equal to the
corresponding bulk jump rates, i.e., $\beta = p, \;\gamma = q$,
and defining reservoir fugacities $z_{r,l}$ by $\alpha = p z_l,
\;\delta = q z_r$, the expression (\ref{steadystate}) reduces to
\be z_k = \frac{(z_l-z_r) \left(\frac{p}{q}\right)^{k-1} + z_r -
z_l \left(\frac{p}{q}\right)^{L-1}}{1-
\left(\frac{p}{q}\right)^{L-1}}\;. \ee This dynamics has a natural
interpretation as coupling the system to boundary reservoirs with
fugacities $z_r,z_l$ respectively. This case has been considered
in \cite{Gros03b} and earlier in \cite{Andj82,deMa84} for $p=q$.
The general solution (\ref{steadystate}) for arbitrary boundary
rates appears to be a new result for the ZRP.
\\
(2) For
\be
\gamma\delta = \alpha\beta \left(\frac{p}{q}\right)^{L-1}
\ee
the current vanishes and the system is in thermal equilibrium.
For the symmetric case $p=q$ this implies constant fugacities $z_k$.\\
(3) For symmetric hopping $p=q=1$ the fugacity profile is
generally linear \bel{symmetric} z_k =
\frac{\alpha+\delta+\alpha\beta(L-1) - (\alpha\beta-\gamma\delta)
(k-1)}{\beta+\gamma+\beta\gamma(L-1)}\;, \ee with a current \be c
=
\frac{\alpha\beta-\gamma\delta}{\beta+\gamma+\beta\gamma(L-1)}\;.
\ee Notice that the linear fugacity profile does not imply a
linear density profile except in
the very special case of non-interacting particles where $z_k = \rho_k$.\\
(4) In the totally asymmetric case, $q=0$, we find \bea
z_k & = & \frac{\alpha}{p+\gamma} \equiv z \mbox{ for } k\neq L\;,\\
z_L & = & \frac{(\alpha+\delta)p+\gamma\delta}{\beta(p+\gamma)}\;.
\eea
The current is given by $c=p z$.\\

In the considerations above we have tacitly assumed that
the local partition function $Z_k$ exists for all $k$. Since for
suitable choices of boundary parameters the local fugacities at
the boundary may take arbitrary values this assumption implies an
infinite radius of convergence of $Z_k = Z(z_k)$ which is not the
case in the models described above. This raises the question
of the long-time behaviour of the ZRP for strong boundary drive,
i.e., rates which drive the boundary fugacities out the radius
of convergence of $Z$. This is studied in detail in the following
sections for the condensation model (\ref{condmodel}).
The rod model (\ref{rodmodel}) can be regarded as a generic ZRP
without bulk condensation transition, but finite radius of
convergence for $Z$. Where appropriate we compare its behaviour
with that of the condensation model.

\section{Condensation model --- Steady state and dynamics near the boundary}
\label{sec:finite}

In this section we consider in some detail the long-time behavior
of the condensation model (\ref{condmodel}). For this model the
local fugacity $z_k$ has to satisfy $z_k\leq1$ in the steady
state. On a ring geometry, the model exhibits a condensation for
$b>2$ at high density. We first analyze the totally asymmetric
case, where particles can only hop to the right. The cases of
partial asymmetry and symmetric hopping are then treated.

\subsection{Totally asymmetric hopping}

For the totally asymmetric case we take $q=\gamma=\delta=0$. For
the normalization of time we set $p=1$. The exact steady-state
solution (\ref{steadystate}) yields \bea
\label{tas1} z_k & = & \alpha \equiv z \mbox{ for } k\neq L\;,\\
\label{tas2} z_L & = & \frac{\alpha}{\beta} \;, \eea and the
current is given by $c=z$. Since for this model $z$ has to satisfy
$z\leq1$, the steady state (\ref{tas1}, \ref{tas2}) is valid only
for $\alpha \leq 1$ and $\beta \geq \alpha$. In this case the
single-site steady-state distribution is given, for large $n$, by
$P^*(\eta_k=n) \sim z^n/n^b$.

We now proceed to discuss the dynamical behavior of the model in
the case that stationary state does not exist. For $\alpha>1$ the
following picture emerges.

\vspace{12pt}\noindent {\bf Site 1:}\\[4mm]
On site~1 particles are deposited randomly with rate $\alpha > 1$
and are removed by hopping to site 2 with rate $1+b/n_1$. Hence
the occupation number performs a simple biased random walk on the
set $n_1$ of positive integers with drift $\alpha - 1 - b/n_1$
which is positive for any $n_1 > b/(\alpha - 1)$. Such a random
walk is non-recurrent and reaches asymptotically the mean velocity
$v=\alpha - 1$. Hence the mean particle number $N_1(t) =
\exval{n_1(t)}$ on site~1 grows asymptotically linearly
\bel{site1} N_1(t) \sim (\alpha - 1) t. \ee

\vspace{12pt}\noindent {\bf Boundary sites $k>1$:}\\[4mm]
We extend the random walk picture (which is strictly valid for site~1)
to site~2. On site~2 particles are injected (by hopping from site
1) with rate $1 + b/n_1$ and are removed with rate $1 + b/n_2$.
Since $n_1$ increases in time on average the input rate approaches
1 and the occupation number at site 2 performs a biased random
walk with hopping rate 1 to the right and rate $1 + b/n_2$ to the
left. Whether this random walk is recurrent depends on $b$. The
asymptotic behavior has been analyzed in \cite{Levi04}. We merely
quote the result: \bel{site2} N_2(t) \sim \left\{ \ba{ll}
t^{1/2} & b < 1 \\
t^{1/2}/\ln{t} & b = 1 \\
t^{1-b/2} & 1 < b < 2 \\
\ln{t} & b = 2 \\
\rho^\ast & b > 2 \ea \right. \ee
The constant
\bel{critdens} \rho^\ast = \frac{1}{b-2} \ee
is the critical
density of the condensation model \cite{Oloa98}. It is approached
with a power law correction $t^{1-b/2}$. By applying this random
walk picture to further neighboring sites, and assuming scaling,
it has been shown that neighboring boundary sites behave
asymptotically in the same fashion \cite{Levi04}. Similar analysis
shows that the square-root increase of the particle density occurs
also for the model (\ref{rodmodel}) for all values of its interaction
parameter $b'$.

\begin{figure}
\centerline{\includegraphics[height=5.5cm]{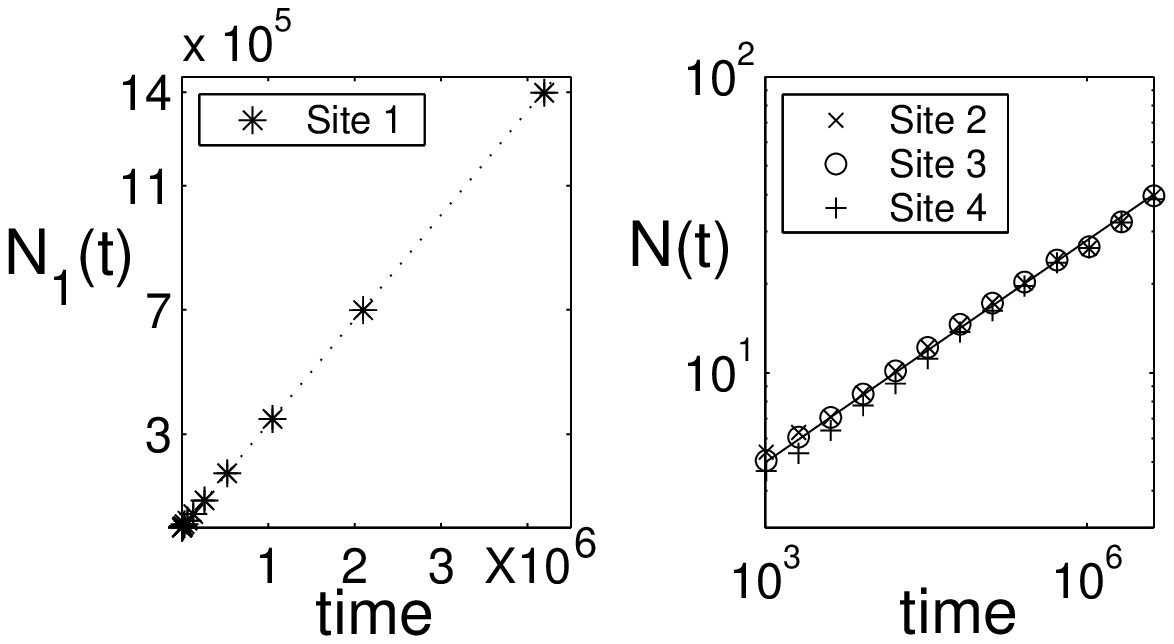}}
\caption{Temporal evolution of local densities, as obtained from
numerical simulations of the totally asymmetric condensation
model, with $b=3/2$, and $\alpha=2$. The solid line corresponds to
the expected growth law $t^{1/4}$, and the dotted line corresponds
to linear growth.} \label{fig:simulation1}
\end{figure}

To test the validity of the random walk picture to sites beyond
$k=2$, we carried out numerical simulations of a totally
asymmetric model with $L=5$. We first note that since the hopping
is totally asymmetric the time evolution of the system up to site
$k$ is independent of what happens at sites to the right of $k$.
In particular, the dynamics on all sites $k<L$ is independent of
$\beta$. Hence, in order to study boundary layers it is sufficient
to simulate very small systems of only a few sites. In
figure~\ref{fig:simulation1} we present the long-time behavior of
the occupation of sites $k=1 - 4$ for $\alpha=2$ and $b=3/2$. It
is readily seen that while site $1$ grows linearly in time, sites
$2-4$ grow with the expected power law $t^{1/4}$.

\vspace{12pt}\noindent {\bf Bulk sites $k\gg 1$:}\\[4mm]
The picture of the simple random walk becomes increasingly
inaccurate as one enters into the bulk of the system, since
injection events onto a site become increasingly correlated in
time. This violates the random walk assumption and makes the
previous analysis invalid in this case. The temporal behavior of
bulk sites $k \gg 1$ can be treated exactly by using a
hydrodynamic approach, which yields the behavior of bulk sites in
the long-time limit. This analysis, carried out in
Sec.~\ref{sec:hydro}, shows a different dynamical behavior in the
bulk. Notice, however, that for any {\it finite} system all ``bulk
sites'' have finite distance from the boundary and hence behave
asymptotically like the boundary sites.

\vspace{12pt}\noindent {\bf Site L:}\\[4mm]
Since the motion of particles on site $k$ is independent of the
motion on sites to their right we expect the bulk result to be
asymptotically valid on all sites up to site $L-1$, i.e., there is
no right boundary layer with yet another set of growth exponents.
On site $L$ the following picture emerges. There is an asymptotic
incoming flux $c=1$ and an exit rate $\beta (1+b/n_L)$. For
$\beta=1$ the boundary site behaves like a bulk site and we obtain
the bulk growth exponent. For $\beta < 1$ the outgoing flux does
not compensate the incoming flux which yields asymptotically
linear growth $\rho_L(t) = (1-\beta) t$. For $\beta > 1$ a finite
stationary chemical potential $z_L = 1/\beta$ is approached.

\begin{figure}
\centerline{\includegraphics[height=5.5cm]{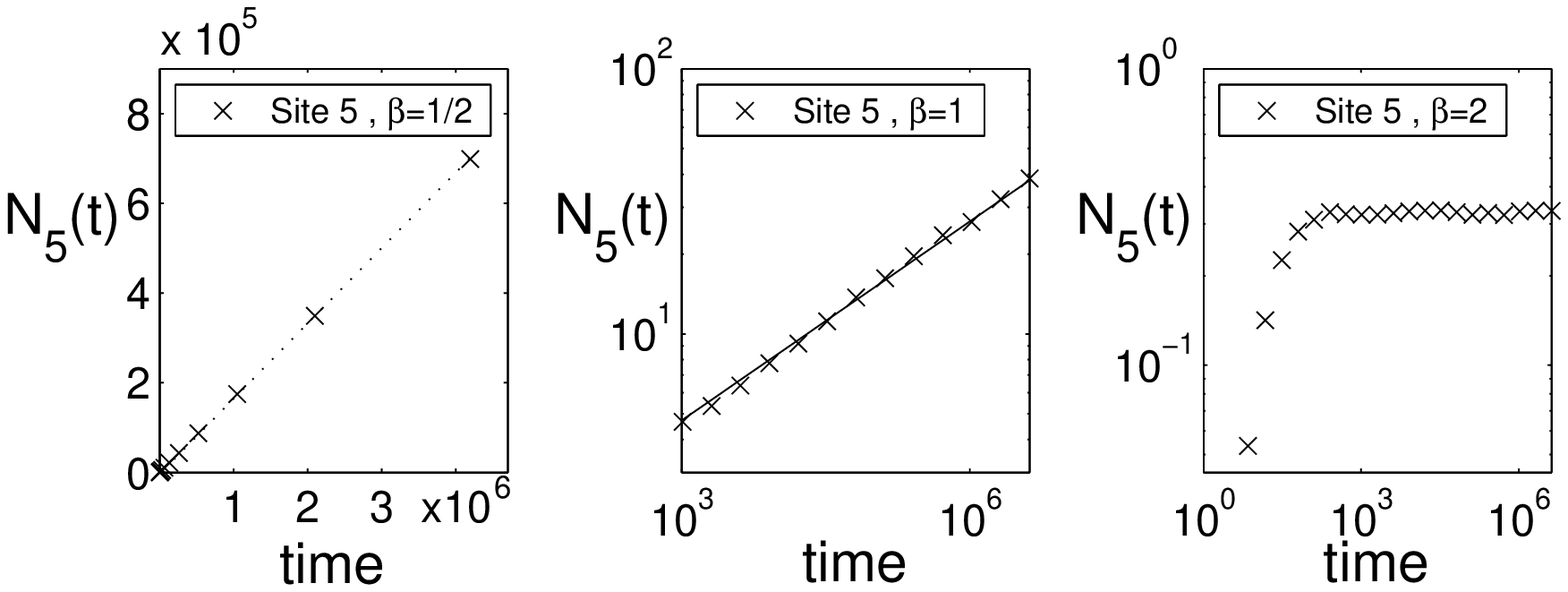}}
\caption{Temporal evolution of local density at the rightmost
site, as obtained from numerical simulations of the totally
asymmetric condensation model, with $b=3/2$, $\alpha=2$ and
$\beta=1/2, 1, 2$. The solid line corresponds to the expected
growth law $t^{1/4}$, and the dotted line corresponds to linear
growth.} \label{fig:simulation2}
\end{figure}

In figure~\ref{fig:simulation2} we present simulation results
for the long-time dynamics of
the occupation of site $k=L=5$. Depending on the value of $\beta$
the occupation number either grows linearly ($\beta<1$), grows
with the same power law as the bulk ($\beta=1$), or approaches a
finite density ($\beta>1$), as expected from the above discussion.

To complete the discussion of the totally asymmetric case
we remark that for a subcritical left boundary $\alpha < 1$ but
supercritical $\beta < \alpha$ the bulk of the
system becomes stationary with fugacity $z_{bulk}=\alpha$. On site
$L$ a condensate develops with linearly increasing particle density
$\rho_L(t) \sim (\alpha-\beta) t$.

\subsection{Partially asymmetric hopping}

We now analyze the dynamical behavior of the partially asymmetric
model in the case where no stationary state exists. We start by
considering the case where the rates at both boundaries are such
that the occupation of the two boundary sites increase linearly
with time. This takes place for $\alpha-\gamma < p-q <
\beta-\delta$. Here sites $k=1$ and $k=L$ act as reservoirs for
the rest of the system. The effective rates at which the
reservoirs exchange particles with the system are
$\alpha_{\mathrm{eff}}=\beta_{\mathrm{eff}}=p$ and
$\gamma_{\mathrm{eff}}=\delta_{\mathrm{eff}}=q$. The
fugacity~(\ref{steadystate}) at sites $k=2,\ldots, L-1$ thus
becomes $z_k=1$. The asymptotic temporal behavior~(\ref{site2})
holds also in this case.

\begin{figure}
\centerline{\includegraphics[scale=0.4]{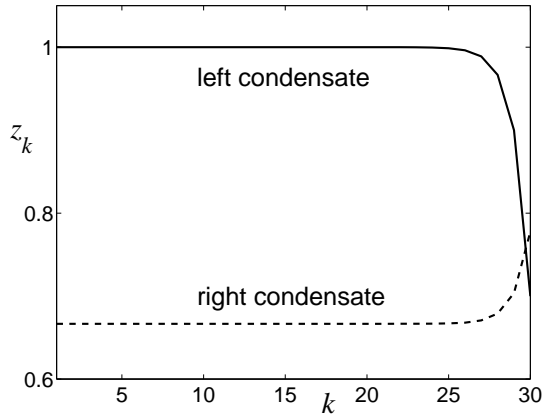}} \caption{The
fugacity profile (\ref{steadystate}) for the partially asymmetric
case, with $b=3/2$, $p=3/4$ and $q=1/4$. Solid line corresponds to
a left condensate ($\alpha=3/4, \beta=1, \gamma=1/4, \delta=1/5$)
and dashed line corresponds to a right condensate ($\alpha=1/2,
\beta=3/4, \gamma=1/4, \delta=1/4$).} \label{fig:zk}
\end{figure}

The picture changes qualitatively if only one boundary fugacity
does not exist. Suppose first that this happens at site 1. In this
case the density on site $1$ will increase linearly with time, as
in the previous totally asymmetric case. This happens when
$\alpha-\gamma > p-q$ and $\beta-\delta>p-q$. As before, site $1$
acts as a reservoir for the rest of the system, with effective
rates $\alpha_{\mathrm{eff}}=p,\; \gamma_{\mathrm{eff}}=q$. Sites
$k=2,\ldots, L-1$ are thus stationary, with the fugacity given by
(\ref{steadystate}). The fugacity at sites away from the right
boundary approaches $1$, with a deviation exponentially small in
the system size (see figure~\ref{fig:zk}). Hence, starting from an
empty initial state one expects algebraic growth of the local
density until, after a long crossover time, stationarity is
reached.

If, on the other hand, the right boundary rates drive site $L$ out
of equilibrium, then the density on site $L$ increases linearly in
time. This happens when  $\alpha-\gamma < p-q$ and $\beta-\delta
<p-q$. Site $L$ acts effectively as a boundary reservoir with
$\beta_{\mathrm{eff}}=p$ and $\delta_{\mathrm{eff}}=q$. As in the
preceding case the system becomes stationary, but with a finite
(independent of system size) deviation of the bulk fugacity from
the critical value $z=1$. Both in the bulk and at the left
boundary the density approaches the finite value dictated by the
left boundary fugacity (see figure~\ref{fig:zk}).

Finally, for $ \beta-\delta < p-q < \alpha-\gamma$ both boundary
sites have finite fugacity, and the system reaches a non-critical
steady state, as given by (\ref{steadystate}).

\subsection{Symmetric hopping}

The analysis of symmetric hopping ($p=q=1$) follows very closely
that of the partially asymmetric case, except that here the
fugacity profile is linear rather than exponential. In particular,
if only one boundary is driven out of equilibrium, a condensate
appears at that boundary with a particle density which increases
linearly with time. This site, say site $1$, acts as a reservoir
with $\alpha_{\mathrm{eff}} = \gamma_{\mathrm{eff}}=1$. The
fugacity profile decreases linearly from $1-O(1/L)$ at the left
boundary to $\delta/\beta+O(1/L)$ at the right boundary, as given
by~(\ref{symmetric}).

When both boundary sites are supercritical they act as reservoirs
with effective rates $\alpha_{\mathrm{eff}} =
\beta_{\mathrm{eff}}=\gamma_{\mathrm{eff}} =
\delta_{\mathrm{eff}}=1$, yielding $z_k=1$ throughout the system.
This is also the case relevant for studying the fluid-mediated
interaction of probe particles in two-species systems
\cite{Levi04}. We expect similar boundary growth laws as in the
asymmetric case. This is because the incoming flux from the
direction of the bulk is compensated by the outgoing flux at the
boundary and one has an effective random picture with the same
rates as above. However, corrections to scaling are expected to be
larger as current correlations are now stronger due to hopping
contributions from the bulk. Moreover, as opposed to the
asymmetric case, in a semi-infinite system both boundaries have
the same growth exponents.

\section{Hydrodynamics --- Exact analysis of dynamics in the bulk}
\label{sec:hydro}

In order to analyze the time evolution of sites far away from the
boundaries we consider the hydrodynamic limit of the ZRP model.
The coarse-grained time evolution of the density profile starting
from a non-stationary initial profile can be determined, by
adapting standard arguments \cite{Kipn99,Spoh91}, from the
continuum limit of the lattice continuity equation. This equation
reads \bel{continuity} \frac{d}{dt}\rho_k = c_{k-1} - c_{k}\;, \ee
with the local current \bel{bulkcurrent} c_{k} = p z_{k} - q
z_{k+1}\;, \ee and $z$ expressed in terms of $\rho$. Together with
an appropriate choice of constant boundary fugacities the solution
is uniquely determined \cite{Gros03b}.

For the driven system one obtains under Eulerian scaling (lattice
constant $a\to 0$, $t \to t/a$, system length fixed) \bel{Euler}
\frac{\partial}{\partial t}\rho(x,t) = - (p-q)
\frac{\partial}{\partial x} z(x,t) + a
(p+q)/2\frac{\partial^2}{\partial x^2} z(x,t) \ee where the
infinitesimal viscosity term serves as regularization and selects
the physical solution of the otherwise ill-posed initial value
problem with fixed boundary fugacities. It takes care of a proper
description of discontinuities which may arise in the form of
shocks or a boundary discontinuity. Indeed, in the large-time
limit the density approaches a constant given by the left boundary
fugacity, with a jump discontinuity at the right boundary
\cite{Gros03b}. This is in agreement with the exact result derived
in the previous subsection. We stress that (\ref{Euler}) provides
an exact description of the density evolution under Eulerian
scaling. It is not a continuum approximation involving a mean
field or other assumption. The simple form of (\ref{Euler})
originates in local stationarity. The absence of noise is due to
Eulerian scaling, i.e.,  the effects of noise appear on finer
scales. For a recent rigorous discussion of the hydrodynamic limit
of stochastic particle systems, see \cite{Kipn99,Saad04}.

Consider a semi-infinite system which is initially empty and has
constant density at the left boundary. This boundary condition
induces a rarefaction wave entering the bulk which can be
constructed using the method of characteristics. The speed $v_0$
of the wave front is given by the zero-density characteristic of
(\ref{Euler}) which is the average speed $v_0=(p-q)w_1$ of a
single particle. In light of the results of the previous sections
we are particularly interested in the case where the left boundary
fugacity is equal to 1. For $b<2$ in the condensation model [and
for any $b'$ in the rod model (\ref{rodmodel})] this corresponds
to an infinite boundary density. On the other hand, for $b>2$ the
corresponding boundary density is $\rho_c$~(\ref{critdens}).
Therefore we are searching for a scaling solution in terms of the
scaling variable $u=x/(v_0t)$ such that $\rho(u) = 0$ for $u \geq
1$ while $\rho=\rho_c$ or $\rho=\infty$ at the left boundary. On
physical grounds the solution has to be continuous as no shock
discontinuities can develop for the initial state (empty lattice)
under consideration. Under these conditions (\ref{Euler}) can be
integrated straightforwardly by setting $a=0$ and one finds the
implicit representation \bel{solutionz} \frac{d \, z}{d\,\rho} =
uw_1\;. \ee This is the solution within the interval $v_1t \leq x
\leq v_0t$. Here \bel{coll} v_z = (p-q) \frac{dz}{d\rho} \ee is
the collective velocity of the lattice gas which plays the role of
the speed of the characteristics for the hydrodynamic equation
(\ref{Euler}). Outside this interval one has $z=0$ for $x\geq
v_0t$ and $z=1$ for $0\leq x\leq v_1t$. We now analyze this
solution in terms of $\rho$.

For the model (\ref{rodmodel}) one has $Z=1/(1-z)^{b'+1}$ which
yields the fugacity-density relation $z=\rho/(b'+1+\rho)$. Hence
\be \rho(u) = (b'+1)\left(\sqrt{\frac{1}{u}} - 1\right). \ee with
$v_0=(p-q)/(b'+1)$ and $v_1=0$ for all $b'$. At any fixed $x$ the
bulk density increases algebraically \be \rho_{bulk}(t) \sim
t^{1/2} \ee for any $b'>0$.

For the condensation model the local partition is the
hypergeometric function \be Z = {}_2F_1 (1,1;1+b;z) \ee which does
not admit an explicit representation of $z$ as a function of
$\rho$. However, as the fugacity is an increasing function in time
approaching 1 we may analyze its asymptotic behavior by expanding
the hypergeometric function around $z=1$ \cite{Gros03a}.

\vspace{12pt}\noindent {\bf $b<2$:}\\[4mm]
Here the left boundary fugacity $z=1$ corresponds to infinite left
boundary density. Moreover $v_1=0$, therefore the solution
(\ref{solutionz}) describes the density profile in the interval $0
\leq u \leq 1$.  This
enables us to consider fixed $x$ and study the long time limit.
In order to analyze the small $u$ behavior (i.e.
$z$ close to 1 and $\rho$ large) one has to distinguish two
domains \cite{Gros03a}. For $b<1$ one has $\rho(z) \propto
z/(1-z)$ for large $\rho$ while for $1<b<2$ one finds $\rho(z)
\propto z/(1-z)^{2-b}$. As a function of $u$ we make the ansatz
$\rho \propto 1/u^\kappa$ for the large $t$ (i.e. small $u$)
asymptotics. Inserting this into the differential equation
(\ref{Euler}) yields a consistent solution with \be \rho_{bulk}(t)
\sim t^\kappa \ee only for \bel{kappa}
\kappa = \left\{ \ba{ll} \frac{1}{2}  & \mbox{ for } b < 1 \\
\frac{2-b}{3-b} & \mbox{ for } 1 < b < 2 \ea \right. . \ee We
conclude that the bulk density increases algebraically with the
universal diffusive exponent $1/2$ for $b<1$ and with a
non-universal $b$-dependent exponent in the range $1 < b < 2$ of
the condensation model. At $b=1,2$ there are logarithmic
corrections which we do not discuss further.

For symmetric hopping one describes the density dynamics under
diffusive scaling $t \to t/a^2$ and obtains
\bel{diffusion}
\frac{\partial}{\partial t}\rho(x,t) =
 p \frac{\partial^2}{\partial x^2} z(x,t)
\ee
which needs no further regularization. Repeating the previous
analysis one finds the same bulk growth exponents as for the
asymmetric hopping model.

\vspace{12pt}\noindent {\bf $b>2$:}\\[4mm]
In the condensation regime the boundary density corresponding to
$z=1$ is $\rho_c$~(\ref{critdens}). Again two different regimes
are found from the asymptotic analysis of the hypergeometric
function. For $b<3$ one has $v_1=0$ and repeating the same
analysis as for $1<b<2$ shows that at fixed $x$ the density
approaches $\rho_c$ with a power law correction with the same
exponent $(2-b)/(3-b)$ as before. This is analogous to the
behavior in the boundary sites analysed in the previous section,
but the exponent is different. For $b>3$ one finds for the
collective velocity \cite{Gros03a} \be v_1 = (p-q)
\frac{(b-3)^2(b-2)^2}{(b-1)^2} > 0. \ee Hence an analysis of the
long-time behaviour at fixed $x$ is not meaningful. A domain with
constant $\rho=\rho_c$ spreads into the system, with a front speed
$v=v_1$. This front is preceded by the rarefaction wave
(\ref{solutionz}) for $v_1 t < x < v_0 t$. This behaviour as a
function of $b$ is unexpected as usually changes in the
rarefaction wave of this type are caused by changing the boundary
density rather than an interaction parameter of the driven system.

We stress that there are two questions that cannot be answered by
the hydrodynamic analysis given above. First we apply Eulerian or
diffusive scaling respectively. This leaves generally open what
happens in any semi-infinite lattice system at finite lattice
distance from the boundaries or in a finite system. Any deviation
from the results given above which decays on lattice scale as one
approaches the bulk cannot be detected within the hydrodynamic
description. Boundary layers, which have been analyzed in the
previous section and found to have an interesting microscopic
structure, would under scaling at best appear as a structureless
boundary discontinuity.

Secondly, the radius of convergence of $Z$ is 1 and forcing the
boundary fugacities $z_1$ or $z_L$ to be larger than 1 implies a
breakdown of the assumption of local stationarity underlying the
hydrodynamic description, at least in the vicinity of the
boundaries. In particular, the hydrodynamic approach is not
applicable for analyzing the condensation regime $b>2$ if the
boundary density exceeds the critical density of the bulk.
However, this regime may be analyzed by the random-walk approach
discussed in the previous section.

\section{Conclusions}
\label{sec:summary}

In this paper we studied the dynamical behavior of the zero-range
process with open boundary conditions for arbitrary bulk and
boundary rates. It is found that for a weak boundary drive the
model reaches a steady state. The exact steady-state distribution
is calculated and is shown to be a product measure characterized
by site-dependent fugacities. In the case of strong drive the
system does not reach a steady-state, and its evolution in time is
calculated. To this end we considered the condensation model with
an initially empty lattice and studied how the local density
evolves in time. As long as the bulk dynamics does not permit
condensation ($b<2$) the growth of the local density is algebraic
in time with exponents that we determined using a random walk
picture for the boundary region, and standard hydrodynamic
description in the bulk. From this analysis we are led to the
conclusion that in the condensation model with $b>2$ (where
condensation in a periodic system sets in above the critical bulk
density $\rho_c=1/(b-2)$) only the boundary sites develop into a
condensate, with a density increasing linearly in time. All bulk
sites become stationary with finite local fugacities determined by
the boundary rates. Somewhat surprisingly the driven and the
symmetric model have the same bulk and boundary growth exponents.
The boundary condensate appears also for $b<2$ when no
condensation transition exists in a periodic system. It is a
general feature of zero-range processes in which $Z$ has a finite
radius of convergence.

For $b>1$ we observe a precursor to the condensation transition in
the sense that the universal diffusive growth for $b<1$ breaks
down. Bulk and boundary growth exponents become different and both
decrease with $b$, i.e. they become non-universal. It is
interesting to note that $b=1$ plays a special role also for the
stationary state on a ring geometry. For $b<1$ the stationary
probability to find any given site empty vanishes as the density
is increased to infinity \cite{Gros03a}, in agreement with
intuition. However, for $b>1$ every site has a finite probability
of being empty, even if the particle density is infinite. Applied
to present scenario this implies the counterintuitive result that
even at very late time, when the average particle density in an
open system tends to infinity on {\it each site}, one still
expects to find a finite fraction of empty sites at any given
moment.

In this context it is also instructive to study the mean first
passage time (MFPT) of the boundary particle density, i.e. the
mean time $\tau$ after which a particle number $N$ has been
reached for the first time at a given site, starting from an empty
site. Using the exact general MFPT expression \cite{Murt89} for
the effective random walk defined above one finds \be \tau =
\frac{1}{b-1} \left[ \left(\ba{c} N+b \\ b+1 \ea \right) -
\left(\ba{c} N+1 \\ 2 \ea \right) \right] \ee where for
non-integer $b$ the factorials are defined by the
$\Gamma$-function. For large $N$ this quantity has the asymptotic
behavior \be \tau \sim \left\{ \ba{ll}
N^2        & (b<1) \\
N^2\ln{N}  & (b=1) \\
N^{1+b}    & (b>1) \ea \right. . \ee Again there is a transition
at $b=1$, with diffusive behavior for $b<1$ and sub-diffusive
exploration of the state space for $b>1$. In the bulk a simple
random picture for the on-site density dynamics is not valid and a
prediction for the bulk MFPT is not possible. For small finite
system size one expects boundary behavior everywhere, but with
increasing corrections to scaling as one moves away from the
boundary. The precise nature of the crossover from the boundary
growth exponent to the hydrodynamic bulk growth exponent remains
an open problem.

\ack G.M.S. thanks the Weizmann Institute for kind hospitality.
The support of the Albert Einstein Minerva Center for Theoretical
Physics, Israel Science Foundation, and Deutsche
Forschungsgemeinschaft (grant Schu827/4), is gratefully
acknowledged.

\end{document}